\documentclass[final,onefignum,onetabnum]{siuro210301}

\usepackage{lipsum}
\usepackage{amsfonts}
\usepackage{graphicx}
\usepackage{epstopdf}
\usepackage{algorithmic}
\usepackage{listings}
\usepackage{xcolor}

\usepackage{float}
\usepackage{amsmath,amssymb}
\usepackage{enumitem}
\usepackage{tikz}
\usetikzlibrary{arrows,decorations.markings}
\usepackage{tabularx}
\usepackage{multirow}
\usepackage{multicol}
\usepackage{tikzit}

\tikzstyle{emptycircle}=[fill=white, draw=black, shape=circle, tikzit shape=circle, inner sep=0pt, minimum size=7mm]
\tikzstyle{output}=[fill=white, draw=black, shape=circle, minimum size=3pt]

\tikzstyle{newstyle}=[draw=black, ->]
\usepackage{float}
\usepackage{tabularx}

\usepackage{array}
\newcolumntype{Z}{>{\centering\let\newline\\\arraybackslash\hspace{0pt}}X}

\usepackage{amsmath, amssymb}
\usepackage{graphicx}
\usepackage{tikz}
\usepackage{subcaption}

\ifpdf
  \DeclareGraphicsExtensions{.eps,.pdf,.png,.jpg}
\else
  \DeclareGraphicsExtensions{.eps}
\fi

\usepackage{enumitem}
\setlist[enumerate]{leftmargin=.5in}
\setlist[itemize]{leftmargin=.5in}

\newsiamremark{remark}{Remark}
\newsiamremark{hypothesis}{Hypothesis}
\crefname{hypothesis}{Hypothesis}{Hypotheses}
\newsiamthm{claim}{Claim}

\headers{Identifiability and observability properties of linear
compartmental models}{Natali Gogishvili}

\title{Database for identifiability properties of linear compartmental models}

\author{Natali Gogishvili\thanks{École polytechnique, Institute Polytechnique de Paris, France 
  (\email{natali.gogishvili@polytechnique.edu}).}}

\dedication{\small\textit{Project advisor: Gleb Pogudin}}

\usepackage{amsopn}

\ifpdf
\hypersetup{
  pdftitle={An Example SIURO Article},
  pdfauthor={D. Doe, P. T. Frank, and J. E. Smith}
}
\fi

\begin{document}

\maketitle

\begin{abstract}
Structural identifiability is an important property of parametric ODE models.
When conducting an experiment and inferring the parameter value from the time-series data, we want to know if the value is globally, locally, or non-identifiable.
Global identifiability of the parameter indicates that there exists only one possible solution to the inference problem, local identifiability suggests that there could be several (but finitely many) possibilities, while non-identifiability implies that there are infinitely many possibilities for the value. 
Having this information is useful since, one would, for example, only perform inferences for the parameters which are identifiable.
Given the current significance and widespread research conducted in this area, we decided to create a database of linear compartment models and their identifiability results.
This facilitates the process of checking theorems and conjectures and drawing conclusions on identifiability.
By only storing models up to symmetries and isomorphisms, we optimize memory efficiency and reduce query time. 
We conclude by applying our database to real problems. 
We tested a conjecture about deleting one leak of the model states in~\cite{Defs2}, and managed to produce a counterexample.
We also compute some interesting statistics related to the identifiability of linear compartment model parameters.

\end{abstract}

\section{Introduction}
For this project we were interested in the structural identifiability of the parameters --- studying if and to what extent can one recover parameter values from the (noise-free) data.
The parameter of a model is globally identifiable if a solution to the parameter inference problem of an experiment is unique, locally identifiable if number of solutions to the inference problem is finite, and non-identifiable otherwise.
For an interesting and important class of models, linear compartment models, 
we created a database containing the identifiability properties for the models up to four compartments. 
We used models up to graph isomorphisms, since isomorphic models have identical identifiability (up to re-numeration). 
We further explain our methods and go through querying the database in this paper. The source code of the project can be found at: \url{https://github.com/Natali124/LinearCompartmentModels}.

We showcase how our database can be used on Conjecture 4.5 stated by E. Gross, H. Harrington, N. Meshkat, and A. Shiu in~\cite{Defs2}. 
Contrary to what was conjectured, we found examples showing that it is possible to improve identifiability by removing a leak in a model, and, thus, we disproved the conjecture. 
This is surprising since one would expect that additional leak would always provide more information about the parameters.
We provide the smallest counterexamples and suggest explanations.

The paper is organized as follows. We explain preliminaries in \cref{sec:preliminaries}, where we define useful definitions and prerequisites for this paper. We talk about the database in \cref{sec:database}, where we explain the structure of the database and offer a short user manual. We have two sections for applications: you can find identifiability statistics in \cref{sec:identifiability_statistics} and counterexample to the conjecture in \cref{sec:counterexample}. We end with conclusions in \cref{sec:conclusions}.

\section{Preliminaries}
\label{sec:preliminaries}

\subsection{Linear Compartment Models}

A linear compartment model is represented by a directed graph and three subsets of vertices for inputs, outputs, and leaks. 
We can represent it as a set of compartments where material transfers from one compartment to another (where nodes are compartments and edges show the flow between them). 
We are allowed to have leaks (leakage of material outside the system of some compartments), inputs (inputting material into the system into some compartments), and outputs (measuring the amount of the material in a compartment). 
The edges, together with inputs and leaks, have scalar parameters associated to them (rate constants). 
These parameters give us information about the rate of flow from one compartment to another. See \cite[Section 2]{Defs2} for further details.

We can transform a linear compartment model into a system of ODEs. 
The transformation goes as follows: For each term/node $x_i$, we write an equation for $x'_i$. 
If there exists an edge from $x_i$ to $x_j$ for some $j$, then we will add $a_{ji}x_i$ to the equation for $x_j'$ and $-a_{ij}x_i$ to the equation for $x_i'$, where $a_{ji}$ is a rate constant of this edge.
If we have a leak in $x_i$, we add $-a_{-1 i}x$ to the equation, where $a_{-1 i}$ is a leak coefficient. 
If there is an input in $x_i$, then we add an external input function $u_i$ to the equation. 
If there is an output in $x_i$, we do not change the equation for $x'_i$, but we add a new equation $y = x_i$ to the system. 
This corresponds to the intuition behind the models as, based on this system, change of material in the compartment depends on how much material it looses and gains over time and with which rate~\cite{Defs2}.
An example of such transformation is given on Figure~\ref{fig:graph}.

\setcounter{table}{0}
\begin{table}[H]
    \centering
    \begin{tabularx}{\linewidth}{Z|Z}
\begin{tikzpicture}
        \node (1) at (1, 1) [circle, draw=black] {0};
        \node (2) at (1, 3) [circle, draw=black] {1};
        \node (3) at (3, 1) [circle, draw=black] {2};
        \node (leak) at (0, 0) [draw=none] {};
        \node (out) at (1, 0) [circle, draw=black] {};
        \node (in) at (3, 2) [draw=none] {};
        \draw[decoration={markings,mark=at position 1 with
    {\arrow[scale=1,>=stealth]{>}}},postaction={decorate}] ([xshift=1mm]1.north) -- ([xshift=1mm]2.south) node[right, pos=0.5, font=\Large] {$a_{10}$};
         \draw[decoration={markings,mark=at position 1 with
    {\arrow[scale=1,>=stealth]{>}}},postaction={decorate}] ([xshift=-1mm]2.south) -- ([xshift=-1mm]1.north) node[left, pos=0.5, font=\Large] {$a_{01}$};
          \draw[decoration={markings,mark=at position 1 with
    {\arrow[scale=1,>=stealth]{>}}},postaction={decorate}] (3) -- (1) node[above, pos=.5, font=\Large] {$a_{02}$};
           \draw[decoration={markings,mark=at position 1 with
    {\arrow[scale=1,>=stealth]{>}}},postaction={decorate}] (1) -- (leak) node[above, sloped, pos=.5, font=\Large] {$a_{-10}$};
           \draw[decoration={markings,mark=at position 1 with
    {\arrow[scale=1,>=stealth]{>}}},postaction={decorate}] (1) -- (out);
            \draw[decoration={markings,mark=at position 1 with
    {\arrow[scale=1,>=stealth]{>}}},postaction={decorate}] (in) -- (3);
\end{tikzpicture}     
      &    
        {\begin{equation*}
        \begin{cases}
          x_0' = \mathbf{-a_{-10}x_0} - \underline{a_{10}x_0} + a_{01}x_1 + a_{02}x_2\\
          x_1' = \underline{a_{10} x_0} - a_{01}x_1\\
          x_2' = - a_{02}x_2 + u_2\\
          y = x_0
        \end{cases}\end{equation*}}
    \end{tabularx}
\captionsetup{name=Figure}
\caption{Example of a linear compartment ODE model and a corresponding system}\label{lincomp_ex}
\label{fig:graph}
\end{table}

\subsection{Parameter identifiability}
Parameter can be globally identifiable, locally identifiable, or non-identifiable.
Globally identifiable denotes that we can determine the parameter value uniquely from the time series data for the outputs. 
Locally identifiable means that we can determine the parameter value only up to finitely many values. 
Finally, non-identifiable parameter can have infinitely many values for the same data.
For further details we refer to~\cite{Defs}.

\Cref{fig:sym_model} shows an example of locally identifiable parameters. The corresponding ODE system for this model would be:
\begin{equation*}
    \begin{cases}
      x_0' = - a_{20}x_0\\
      x_1' = - a_{21}x_1\\
      x_2' = a_{20}x_0 + a_{21}x_1\\
      y = x_2
    \end{cases}
\end{equation*}
Sometimes it is easy to understand why parameters are locally identifiable. In this example, we see symmetry along the observed edge. Therefore, if we exchange the value of $a_{21}$ with the value of $a_{20}$ and $x_1(0)$ with $x_0(0)$ (the initial conditions) the solution for $x_2(t)$ will stay the same. 
Therefore, there is no way to determine the reaction rates based on the time-series for $x_2$ independently of the quality of the data.

\setcounter{figure}{1}
\begin{figure}[h!]
\begin{center}
    \includegraphics[scale=0.7]{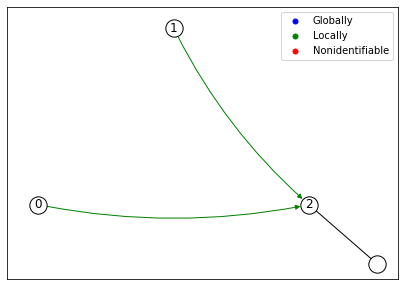}
    \caption{Example of the model which is locally identifiable}
    \label{fig:sym_model}
\end{center}
\end{figure}

\section{The database}
\label{sec:database}

We created a database of linear compartment models with up to (including) 4 vertices to easily check identifiability conjectures and verify theorems up to certain number of vertices~\cite{SC}. In total database consists of $73,416$ models (more precisely, isomorphism classes of models): of 2 nodes --- $32$, of 3 nodes --- $920$, and of 4 nodes --- $72,464$. In fact, it supports many more queries since we consider models up to graph isomorphism.

As mentioned, we needed to create a database of models together with identifiability results. There were several challenges to consider while doing this. Most importantly, we did not want too many models to be in the database as this would cause increase in computation time of identifiability while generating the models as well as the increase of query time. Because of this, we decided to generate models up to graph isomorphism. Isomorphic models have exactly the same properties as they only differ by labeling of the nodes. Therefore, considering only one model from all isomorphic models saves a lot of time and memory. Additionally, we only chose to consider models where all vertices reach at least one output as otherwise, we could replace this part with a leak. Lastly, graph is required to be (not necessarily strongly) connected since otherwise such a model is simply a union of two smaller models.

The easiest way to query the database is to use the class \texttt{Data} from the file {\sc IdentifiabilityResults.py}. 
Constructor takes as an argument the name of the directory where the results are located. 
For example, below you will find a walk-through example which can be run from the root directory of the repository.

\begin{lstlisting}{python}

# Following prepares an object D of Data class with all our data
D = Data('results')

\end{lstlisting}

Now we could use the following code to find all models with 2 nodes and 1 leak which have at least one globally identifiable parameter.

\begin{lstlisting}{python}

def condition(m):
    return len(m.leaks) == 1 and len(m.graph) == 2

filtered_models = D.filterby(condition)

for model, result in filtered_models.items():
    found = False
    for parameter, identifiability in result.items():
        if identifiability == 'globally':
            found = True
            break
    if found:
        print(f'Model: {model}')
        print(f'Result: {result}\n')

\end{lstlisting}

This will print the following to the standard output:
\begin{lstlisting}{bash}

Model: Graph: [set(), {0}], Inputs: {0}, Outputs: {0}, Leaks: {0}
Result: {'(0, -1)': 'globally', '(1, 0)': 'globally'}

Model: Graph: [set(), {0}], Inputs: {1}, Outputs: {0}, Leaks: {0}
Result: {'(0, -1)': 'globally', '(1, 0)': 'globally'}

Model: Graph: [set(), {0}], Inputs: {1}, Outputs: {0}, Leaks: {1}
Result: {'(1, -1)': 'globally', '(1, 0)': 'globally'}

Model: Graph: [{1}, {0}], Inputs: {0}, Outputs: {0}, Leaks: {0}
Result: {'(0, -1)': 'globally', '(0, 1)': 'globally', '(1, 0)': 'globally'}

Model: Graph: [{1}, {0}], Inputs: {0}, Outputs: {0}, Leaks: {1}
Result: {'(1, -1)': 'globally', '(0, 1)': 'globally', '(1, 0)': 'globally'}

Model: Graph: [{1}, {0}], Inputs: {0}, Outputs: {1}, Leaks: {0}
Result: {'(0, -1)': 'locally', '(0, 1)': 'globally', '(1, 0)': 'locally'}

Model: Graph: [{1}, {0}], Inputs: {0}, Outputs: {1}, Leaks: {1}
Result: {'(1, -1)': 'globally', '(0, 1)': 'globally', '(1, 0)': 'globally'}

Model: Graph: [set(), {0}], Inputs: {0, 1}, Outputs: {0}, Leaks: {0}
Result: {'(0, -1)': 'globally', '(1, 0)': 'globally'}

Model: Graph: [set(), {0}], Inputs: {0, 1}, Outputs: {0}, Leaks: {1}
Result: {'(1, -1)': 'globally', '(1, 0)': 'globally'}

Model: Graph: [{1}, {0}], Inputs: {0, 1}, Outputs: {0}, Leaks: {0}
Result: {'(0, -1)': 'globally', '(0, 1)': 'globally', '(1, 0)': 'globally'}

Model: Graph: [{1}, {0}], Inputs: {0, 1}, Outputs: {0}, Leaks: {1}
Result: {'(1, -1)': 'globally', '(0, 1)': 'globally', '(1, 0)': 'globally'}

\end{lstlisting}

Also, we could use any model with up to 4 nodes as a key:

\begin{lstlisting}{python}

model1 = LinearCompartmentModel([[1], [0]], [0, 1], [0], [1])
print(model1)
print(D[model1])

model2 = LinearCompartmentModel([[0], [1]], [1, 0], [1], [0])
print(model2)
print(D[model2])

\end{lstlisting}

This gives the following output:

\begin{lstlisting}{bash}

Graph: [{1}, {0}], Inputs: {0, 1}, Outputs: {0}, Leaks: {1}
{'(1, -1)': 'globally', '(0, 1)': 'globally', '(1, 0)': 'globally'}

Graph: [{0}, {1}], Inputs: {0, 1}, Outputs: {1}, Leaks: {0}
{'(0, -1)': 'globally', '(1, 0)': 'globally', '(0, 1)': 'globally'}

\end{lstlisting}

We could also directly iterate over (model, result) pairs to achieve the same result as above:

\begin{lstlisting}{Python}

for model, result in D:
    if len(model.leaks) == 1 and len(model.graph) == 2:
        found = False
        for parameter, identifiability in result.items():
            if identifiability == 'globally':
                found = True
                break
        if found:
            print(f'Model: {model}')
            print(f'Result: {result}')
            print()

\end{lstlisting}

We used the class \texttt{LinearCompartmentModel} to represent the models. 
It has many important functions including \texttt{generating\_model\_isomorphisms} (first generating graph isomorphisms and reshuffling inputs, outputs, and leaks accordingly) and checking strongly connectedness for the conjecture we mentioned. The file {\sc GeneratingModels.py} contains the function \texttt{generate\_models}, which created models with certain number of inputs, outputs, and leaks, and is one of the key components for creating the database. For the second part, assessing the results of identifiability, we used \texttt{assess\_identifiability} function from the Julia library {\sc StructuralIdentifiability.jl} \cite{CodeWeb,structidjl}.

For more detailed instructions on how to use the database, consult the README.md file of the repository.

\section{Application: identifiability statistics}
\label{sec:identifiability_statistics}

We define non-identifiable \emph{models} to be models which have at least one non-identifiable parameter, locally-identifiable \emph{models} to be models which are not non-identifiable and have at least one locally identifiable parameter, and identifiable \emph{models} to be models which have all globally-identifiable parameters.

\begin{figure}[H]
  \centering
  \begin{subfigure}{0.48\textwidth}
    \centering
    \includegraphics[width=\linewidth]{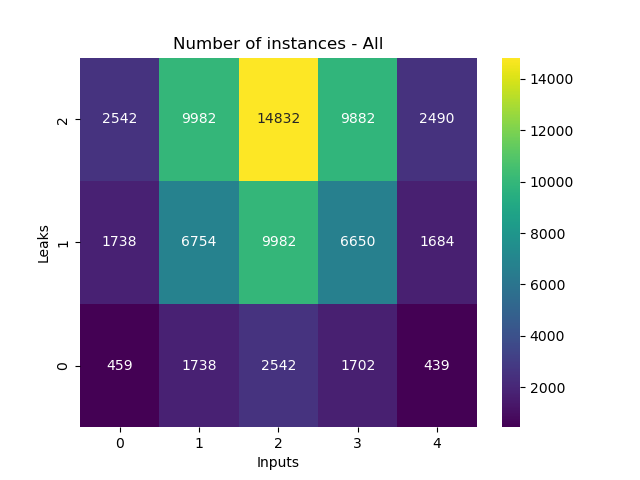}
    \caption{All models.}
    \label{fig:heatall}
  \end{subfigure}
  \hfill
  \begin{subfigure}{0.48\textwidth}
    \centering
    \includegraphics[width=\linewidth]{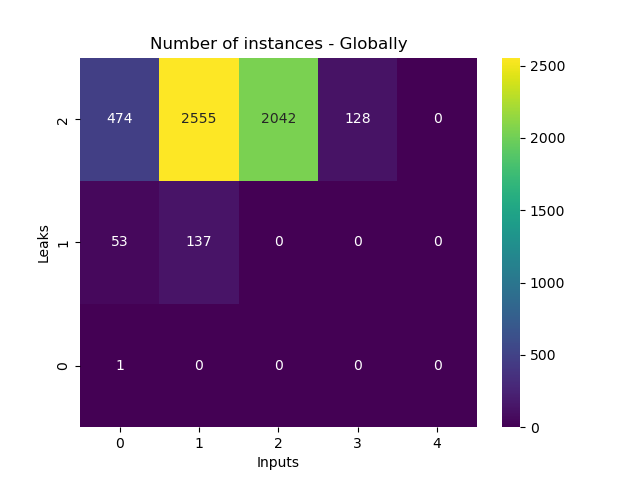}
    \caption{Only globally identifiable models.}
    \label{fig:heatglobally}
  \end{subfigure}

  \begin{subfigure}{0.48\textwidth}
    \centering
    \includegraphics[width=\linewidth]{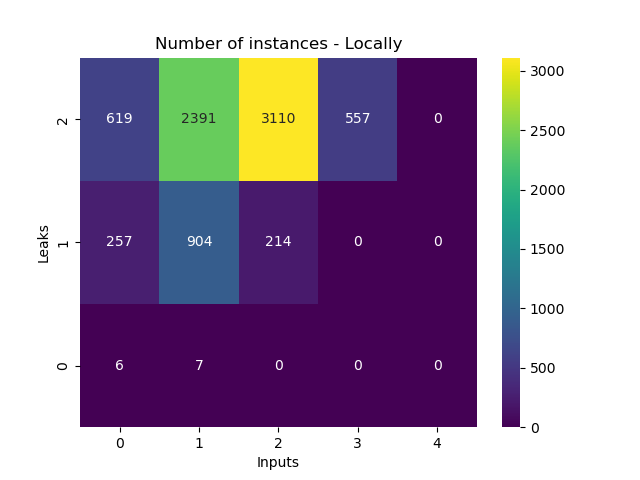}
    \caption{Only locally identifiable models.}
    \label{fig:heatlocally}
  \end{subfigure}
  \hfill
  \begin{subfigure}{0.48\textwidth}
    \centering
    \includegraphics[width=\linewidth]{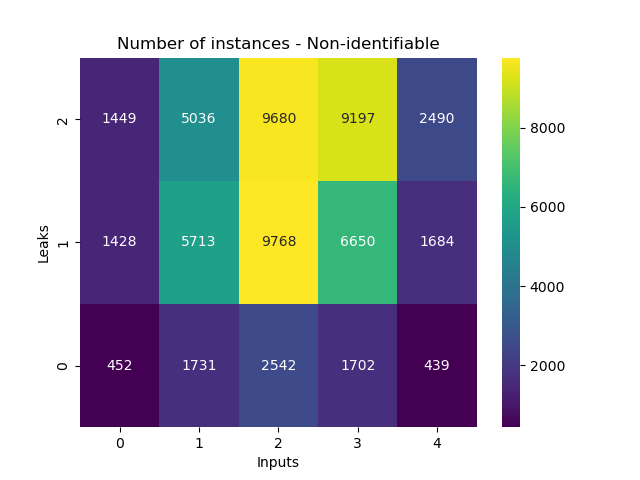}
    \caption{Only non-identifiable models.}
    \label{fig:heatnon}
  \end{subfigure}

  \caption{Heat-maps of results. Number of inputs over number of leaks.}
  \label{fig:fullHeatmaps}
\end{figure}

Below we summarize identifiability statistics over number of nodes, leaks, and inputs (\Cref{fig:combined}). We also suggest heat-plots displaying identifiability considering both leaks and number of inputs (\Cref{fig:fullHeatmaps}). From all the models, 5,390 are globally identifiable, 8,065 are locally identifiable, and 59,961 are non-identifiable.

In the tables displayed we can observe that with the increase of number of leaks, proportion of non-identifiable models generally increases. Opposite is the case for inputs. This is intuitive since generally, adding more leaks to the system increases uncertainty, while increasing number of inputs gives more information about the system. Increased number of non-identifiable parameters when increasing both number of inputs and number of outputs is because the complexity of the models with which more leak/input models are associated is higher.

\begin{figure}[htbp]
    \centering
    \begin{subfigure}{\textwidth}
        \centering
        \begin{tabular}{|c|c|c|c|}
            \hline
            \# vertices & Globally & Locally & Non-Identifiable \\
            \hline
            2 & 19 & 3 & 10 \\
            3 & 228 & 137 & 555 \\
            4 & 5143 & 7925 & 59396 \\
            \hline
        \end{tabular}
        \caption{Identifiability of the models with number of nodes.}
        \label{tab:mytable}
    \end{subfigure}%
    \hspace{1cm}
    \begin{subfigure}{\textwidth}
        \centering
        \begin{tabular}{|c|c|c|c|}
            \hline
            \# leaks & Globally & Locally & Non-Identifiable \\
            \hline
            0 & 528 & 882 & 3329 \\
            1 & 2692 & 3302 & 12480 \\
            2 & 2042 & 3324 & 21990 \\
            3 & 128 & 557 & 17549 \\
            4 & 0 & 0 & 4613 \\
            \hline
        \end{tabular}
        \caption{Identifiability of the models with number of leaks.}
        \label{tab:mytable2}
    \end{subfigure}%
    \hspace{1cm}
    \begin{subfigure}{\textwidth}
        \centering
        \begin{tabular}{|c|c|c|c|}
            \hline
            \# inputs & Globally & Locally & Non-Identifiable \\
            \hline
            0 & 1 & 13 & 6866 \\
            1 & 190 & 1375 & 25243 \\
            2 & 5199 & 6677 & 27852 \\
            \hline
        \end{tabular}
        \caption{Identifiability of the models with number of inputs.}
        \label{tab:mytable3}
    \end{subfigure}
    \caption{Identifiability statistics over number of nodes, leaks, and inputs.}
    \label{fig:combined}
\end{figure}

\section{Application: counterexample to the conjecture}
\label{sec:counterexample}
We got specifically interested in Conjecture 4.5 (deleting one leak) from the paper \emph{Linear Compartmental Models: Input-Output Equations and Operations That Preserve Identifiability} by E. Gross, H. Harrington, N.Meshkat, and A. Shiu~\cite{Conj}.

The conjecture states that if the graph of a model is strongly connected,  the model has at least one input and exactly one leak, if it is locally (or globally) identifiable, then after removing the leak, it will not become non-identifiable. 
We use our database to disprove the conjecture and find the smallest counterexamples.

For this experiment, we created a function \texttt{check4\_5} as a function of \texttt{Data} class which returns a list of all models (up to 3 nodes) with at least one input and exactly one leak which were globally or locally identifiable before removing the leak and non-identifiable after removing the leak. The list was not empty, which disproved the conjecture.

We decided to check those models by hand using the same \texttt{assess\_identifiability} function from \texttt{StructurlaIdentifiability.jl}~\cite{CodeWeb} to avoid possible mistakes. 
We also used a web-based \texttt{Structural Identifiability Toolbox}\footnote{\url{https://maple.cloud/app/6509768948056064/Structural+Identifiability+Toolbox}}  \cite{SIAN} for assessing identifiability, which uses completely different algorithm than \texttt{assess\_identifiability}. 
This all gave us confidence to say that conjecture is disproved.

One interesting result that we obtained was that in the list returned, all models were initially globally identifiable everywhere even though we were also allowing locally identifiable parameters. Similarly, we did not get locally identifiable parameters in the results after removing the leak (they were either non identifiable everywhere or non identifiable and globally identifiable). We also got instances where globally identifiable parameters became locally identifiable after removing the leak, but those cases are less interesting for us.

Here is listed all of the models we got in the list. We tested on models with up to (including) 3 vertices to get minimal counter-examples.
Graphs are represented as adjacency lists (a list of lists where the $i$-th list contains the vertices reachable from the $i$-th node by an edge):

\begin{itemize}
    \item Graph: [[1], [0, 2], [0, 1]], Inputs: [0, 1], Outputs: [1], Leaks: [0]
    \item Graph: [[1], [0, 2], [0, 1]], Inputs: [0, 1], Outputs: [2], Leaks: [0]
    \item Graph: [[1], [0, 2], [0, 1]], Inputs: [0, 2], Outputs: [2], Leaks: [0]
    \item Graph: [[1, 2], [0, 2], [0, 1]], Inputs: [0, 1], Outputs: [0], Leaks: [1]
    \item Graph: [[1, 2], [0, 2], [0, 1]], Inputs: [0, 1], Outputs: [2], Leaks: [0]
\end{itemize}

\begin{figure}[h]
    \centering
    \begin{subfigure}{0.49\textwidth}
        \centering
        \includegraphics[width=\textwidth]{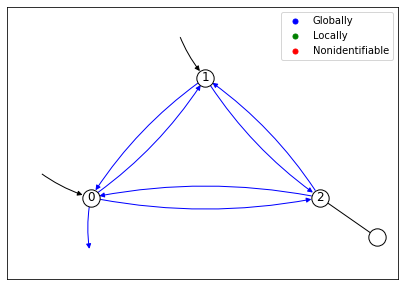}
        \caption{Graph: [[1, 2], [0, 2], [0, 1]], Inputs: [0, 1], Outputs: [2], Leaks: []}
        \label{fig:ex1a}
    \end{subfigure}
    \begin{subfigure}{0.49\textwidth}
        \centering
        \includegraphics[width=\textwidth]{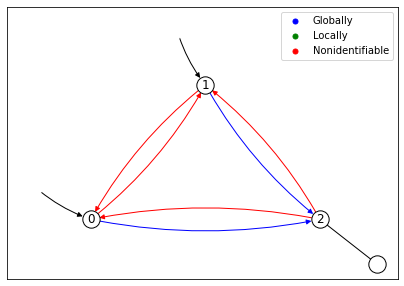}
        \caption{Graph: [[1, 2], [0, 2], [0, 1]], Inputs: [0, 1], Outputs: [2], Leaks: []}
        \label{fig:ex1b}
    \end{subfigure}
    \caption{Illustration of one of the models before and after removing the leak.}
    \label{fig:comb}
\end{figure}

As we can see there are no obvious symmetries in the graphs except of the last one (\Cref{fig:ex1b}). In the last graph we can see that after removing one node we are left with two globally identifiable nodes even though the model is symmetric. By further experiments conducted under (including) 3 nodes, we observed that when there is an edge from one node to another, if second node directly leads to an output and the first node has an input, the edge is always globally identifiable. This might explain result on \Cref{fig:comb} as well.

\begin{figure}[h]
    \centering
    \begin{subfigure}{0.49\textwidth}
        \centering
        \includegraphics[width=\textwidth]{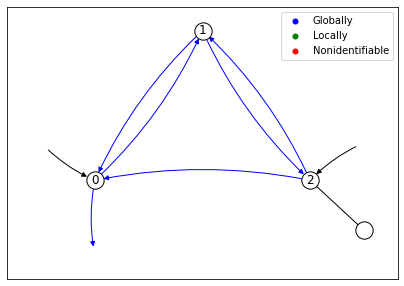}
        \caption{[[1], [0, 2], [0, 1]], Inputs: [0, 2], Outputs: [2], Leaks: [0]}
        \label{fig:ex2a}
    \end{subfigure}
    \begin{subfigure}{0.49\textwidth}
        \centering
        \includegraphics[width=\textwidth]{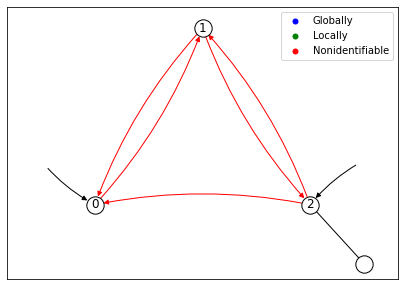}
        \caption{[[1], [0, 2], [0, 1]], Inputs: [0, 2], Outputs: [2], Leaks: []}
        \label{fig:ex2b}
    \end{subfigure}
    \caption{Illustration of one of the models before and after removing the leak.}
    \label{fig:lastmodel}
\end{figure}

The most interesting example is the third model. Before removing the leak all parameters are globally identifiable. After removing the leak all parameters become non-identifiable. We show this example on \Cref{fig:lastmodel}.

\section{Conclusions}
\label{sec:conclusions}

To sum up, during the project we developed a database of linear compartment models with their identifiability results in a way which is suitable for many applications (e.g. verification of conjectures and theorems). We used this database to disprove an important conjecture and showed wrong the intuitive understanding of the leaks as a source of uncertainty.

\section{Acknowledgements}
I want to thank Prof. Gleb Pogudin for supervising me through this project. He had a great contribution towards the success of this project and provided me with invaluable advice and directions. Also, special thanks to Prof. Anne Shiu, one of the authors of the paper from which we disproved the conjecture, for reviewing the results of this paper, expressing interest, and sharing it with her colleagues.
The work was supported by the French ANR-22-CE48-0008 OCCAM project.

\bibliographystyle{siamplain}
\bibliography{references}

\end{document}